# Data-efficient linear hyperspectral imaging through risk-guided co-design

Shenzhan Hong[1,2,3,†], Shuxing Yang[1,2,3,†], Taojun Ying[1,2,3], Fujia Chen[1,2,3,4,*], Zongyin Yang[1,2,3,*], Hongsheng Chen[1,2,3], and Yihao Yang[1,2,3,*]

[1] *State Key Laboratory of Extreme Photonics and Instrumentation, ZJU-Hangzhou Global Scientific and Technological Innovation Center, Zhejiang University, Hangzhou 310027, China.*
[2] *The Electromagnetics Academy at Zhejiang University, College of Information Science and Electronic Engineering, Zhejiang University, Hangzhou 310027, China.*
[3] *Zhejiang Key Lab. of Intelligent Electromagnetic Control and Advanced Electronic Integration, Jinhua Institute of Zhejiang University, Zhejiang University, Jinhua 321099, China.*
[4] *International Research Center for Information Science and Electronic Engineering, Zhejiang University, Haining 314400, China.*

[†]*These authors contributed equally to this work.*
**Corresponding authors. Email: yangyihao@zju.edu.cn (Y. Y.); fujiachen@zju.edu.cn (F. C.); yangzongyin@zju.edu.cn (Z. Y.)*

**Abstract**

Computational hyperspectral imaging reconstructs spatial–spectral information from limited optical measurements. Linear reconstructors are efficient and interpretable, but often underperform deep nonlinear models, which typically require substantial training data and computational resources. Here we present a theory-guided encoding–reconstruction co-design framework that achieves, to our knowledge, state-of-the-art linear reconstruction performance. We establish a theory for the exact five-term decomposition of reconstruction risk and introduce a method for diagnosing the statistical-coverage lower bound within a fixed principal component analysis envelope, which together inform the optimized design of hyperspectral imaging systems. The framework applies across coded-aperture snapshot spectral imaging (CASSI), diffractive optical element imaging and spectral filter array imaging, achieving 34.76 dB mean PSNR in single-shot CASSI, outperforming a deep baseline by 9.43 dB with only five source scenes, and matching empirical cross-domain risk within 0.2%. Our results establish a data-efficient and theoretically grounded route to high-performance linear hyperspectral reconstruction, while providing a quantitative basis for identifying whether reconstruction is limited by optical encoding, reconstruction design or statistical coverage.

## Introduction

Computational hyperspectral imaging recovers high-dimensional spatial and spectral information from limited optical measurements[1–3], enabling applications ranging from remote sensing to biomedical imaging and material identification[4–7]. Conventional linear reconstruction methods are computationally efficient and interpretable, but often struggle to achieve high accuracy under strongly compressed measurements[8]. Deep learning has substantially improved reconstruction performance through nonlinear models, including approaches that jointly optimize optical encoding and computational reconstruction[9–11]. However, these improvements typically come at the cost of substantial training data and computational resources, with performance often deteriorating when training data are scarce. Achieving the accuracy of deep nonlinear reconstruction while retaining the simplicity, efficiency and low data requirements of linear methods remains a key challenge.

Here we introduce a theory-guided encoding–reconstruction co-design framework that achieves, to our knowledge, state-of-the-art linear reconstruction performance in single-shot coded-aperture snapshot spectral imaging (CASSI). We first derive an exact five-term decomposition of reconstruction risk and establish a statistical-coverage lower bound for reconstruction within a fixed principal component analysis (PCA) envelope. Guided by this theory, we jointly optimize the optical encoder and its matched linear minimum mean-square error (LMMSE) reconstructor using a common statistical prior. The resulting linear reconstructor is obtained in closed form, requiring no GPU training while approaching the accuracy of high-performance deep nonlinear methods.

We validate the framework across CASSI, diffractive optical element (DOE) imaging and spectral filter array (SFA) imaging. In single-shot CASSI, our linear reconstructor achieves a mean PSNR of 34.76 dB, approaching that of a high-performance deep reconstructor, and outperforms the same deep baseline by 9.43 dB when only five source scenes are available. Cross-domain experiments show that the analytical risk agrees with empirical reconstruction error within 0.2%, and further reveal that source-data coverage can dominate over encoder optimization in reducing reconstruction risk. These results establish a theory-guided route to high-performance, data-efficient linear hyperspectral reconstruction, while providing a quantitative framework for identifying whether reconstruction performance is limited by optical encoding, reconstruction design or statistical coverage.

## Results

### Theory of exact reconstruction-risk decomposition

We consider an HSI system described by the following linear measurement model:

$$y = H_\theta x + n, \quad \theta \in \Theta \tag{1}$$

where $x$ denotes the hyperspectral signal, $y$ the finite-dimensional measurements, $H_\theta$ the sensing operator determined by the physical encoding $\theta$, $\Theta$ the set of physically realizable parameters, and $n$ the measurement noise. In this formulation, different imaging systems are characterized by their sensing operators and physical constraints.

We first characterize scene statistics using a source-domain envelope defined by principal component analysis (PCA). Let $\mu_s$ denote the source-domain sample mean, $B$ the principal-component basis obtained from the centred source-domain samples, $q$ the out-of-envelope residual, and $a$ the low-dimensional coefficient vector. A source-domain hyperspectral sample can then be expressed as $X_s = \mu_s + Ba + q$. In practice, the prior is estimated from source-domain data, while the target scenes may have different statistics. To characterize this mismatch, we perform PCA on the centred source-domain hyperspectral samples and decompose their covariance as $\Sigma_s = \mathbb{E}_s[(X_s - \mu_s)(X_s - \mu_s)^T] = B\Lambda_{\text{env}}B^T + \Gamma_q$, where $\Sigma_s$ denotes the covariance matrix of the source-domain hyperspectral data, $\Lambda_{\text{env}}$ the diagonal matrix of eigenvalues associated with the principal components, and $\Gamma_q$ the covariance of the low-energy out-of-envelope residual, which may contain sensor noise and perturbations. Separating this residual improves numerical stability and prevents limited measurement resources from being overallocated to weak directions. We therefore define the PCA envelope using the principal components that explain 99.5% of the cumulative variance and use $B\Lambda_{\text{env}}B^T$ as the encoder-design prior. We examine how this threshold affects reconstruction performance in Supplementary Note 5.3 (Supplementary Table 4 and Supplementary Fig. 4). Fig. 1a shows that the envelope dimension increases with the number of source-domain scenes, reflecting broader coverage of source-domain variation. The covariance matrix of the target-domain scenes can then be defined as $\Sigma_t = \mathbb{E}_t[(x - \mu_t)(x - \mu_t)^T] = U_t\Lambda_t U_t^T$, where $\Sigma_\text{t}$ denotes the covariance matrix of the target-domain hyperspectral data, $\mu_\text{t}$ the target-domain mean, and $U_\text{t}$ and $\Lambda_\text{t}$ the principal-component basis and corresponding eigenvalues, respectively. The relationship between the two envelopes is illustrated in Fig. 1b.

Under the bias-corrected linear measurement model in Eq. (1), we assume conditionally zero-mean noise with covariance $R_n$ (Methods). We consider an affine linear reconstructor of the form $\hat{x} = \mu_s + W(y - H_\theta \mu_s)$, where $\hat{x}$ denotes the reconstructed hyperspectral result, and $W$ the linear reconstructor. Combining this expression with the forward model gives the reconstruction error $e = x - \hat{x} = (I - WH_\theta)(x - \mu_s) - Wn$. Defining $\Delta\mu = \mu_t - \mu_s$, the spectral reconstruction risk can be further derived as

$$\mathcal{E}(W, H_\theta; \Sigma_t, \Delta\mu, R_n) = \mathbb{E}_t\left[\|e\|_2^2\right] = \mathrm{tr}\left[(I - WH_\theta)\Sigma_t(I - WH_\theta)^T\right] + \mathrm{tr}\left(WR_nW^T\right) + \|(I - WH_\theta)\Delta\mu\|_2^2 \quad (2)$$

Using the source-domain PCA basis introduced above, we define the orthogonal projector onto the source-domain PCA envelope as $P_B = BB^T$, and the projector onto its orthogonal complement as $Q = I - P_B$. Substituting these projectors into the target-domain covariance term in Eq. (2) and expanding yields an exact decomposition of the reconstruction risk into five analytical components

$$\mathcal{E}(W, H_\theta; \Sigma_t, \mu_t, \mu_s, R_n) = T_{in} + T_{out} + T_{cpl} + T_{noise} + T_{mean} \quad (3)$$

with

$$\begin{aligned}
T_{in} &= \mathrm{tr}\left[(I - WH_\theta)P_B\Sigma_t P_B(I - WH_\theta)^T\right] \\
T_{out} &= \mathrm{tr}\left[(I - WH_\theta)Q\Sigma_t Q(I - WH_\theta)^T\right] \\
T_{cpl} &= \mathrm{tr}\left[(I - WH_\theta)(P_B\Sigma_t Q + Q\Sigma_t P_B)(I - WH_\theta)^T\right] \\
T_{noise} &= \mathrm{tr}\left(WR_nW^T\right) \\
T_{mean} &= \|(I - WH_\theta)(\mu_t - \mu_s)\|_2^2
\end{aligned}$$

where $T_{in}$ denotes the reconstruction-risk contribution from the target-domain component within the source-domain PCA envelope. $T_{out}$ denotes the risk associated with target-domain components that lie outside or are misrepresented by the source-domain PCA envelope. $T_{cpl}$ is the cross-coupling term between the in- and out-of-envelope components and can be either positive or negative. $T_{noise}$ is the noise-propagation risk induced by measurement noise passing through the reconstructor, and $T_{mean}$ denotes the source-target mean-shift term. The exactness of the five-term decomposition is independently verified by Monte Carlo experiments in Supplementary Note 3, with the analytical and numerical results reported in Supplementary Fig. 1 and Supplementary Table 1. Eq. (3) provides a complete decomposition of the target-domain spectral reconstruction risk and applies to any finite-dimensional parameterized linear sensing operator and affine linear reconstructor.

**Method of PCA-envelope-guided risk diagnosis and co-design**

Estimating a full covariance matrix from limited source data can be unreliable[12,13], so we use a low-rank PCA model. At a fixed rank, PCA minimizes the mean-squared projection error of the source data. Our reconstructor estimates only the retained PCA coefficients, so reconstructed deviations from the source mean remain within the PCA envelope. Under this constraint, the reconstruction risk separates as follows:

$$\begin{aligned}
&\mathcal{E}(W, H_\theta; \Sigma_t, \mu_t, \mu_s, R_n) \\
&= \underbrace{\mathrm{tr}(Q\Sigma_t) + \|Q(\mu_t - \mu_s)\|_2^2}_{R_{stat}} + \underbrace{\mathrm{tr}\left[(P_B - WH_\theta)\Sigma_t(P_B - WH_\theta)^T\right] + \mathrm{tr}(WR_nW^T) + \|(P_B - WH_\theta)(\mu_t - \mu_s)\|_2^2}_{R_{design}} \quad (4)
\end{aligned}$$

Eq. (4) separates the reconstruction risk into a statistical-coverage term, $R_{stat}$, and a design-dependent term, $R_{design}$. For a fixed PCA envelope, $R_{stat}$ captures the target-domain covariance energy and mean-shift error outside the envelope. This term is independent of the encoder and reconstructor and sets a lower bound on the risk of PCA-envelope-constrained reconstruction. $R_{\text{design}}$ accounts for the remaining in-envelope estimation error, noise propagation and in-envelope mean-shift error, which may be reduced through co-design. The same separation holds for other fixed orthogonal projection subspaces, as shown in Supplementary Note 1.1.

More directly, let $R_0$ denote the reconstruction risk of a current feasible design and $R^*$ the minimum risk attainable within the same physically realizable design space and PCA-envelope-constrained reconstruction class, with the source-domain PCA envelope fixed. Then:

$$0 \le \frac{R_0 - R^*}{R_0} \le 1 - \frac{R_{\text{stat}}}{R_0} \tag{5}$$

Eq. (5) bounds the possible risk reduction from further encoder–reconstructor co-design at a fixed PCA envelope. It therefore provides a quantitative way to assess whether further improvement should focus on statistical coverage or system optimization.

Using the source-domain PCA prior, we jointly determine the physical encoder and its matched linear minimum mean-square error (LMMSE) reconstructor. The encoder is optimized by minimizing the trace of the PCA coefficient-estimation error covariance:

$$\begin{aligned} C_{\text{LMMSE}}(\theta) &= \left[\Lambda_{\text{env}}^{-1} + B^{\text{T}} H_{\theta}^{\text{T}} R_n^{-1} H_{\theta} B\right]^{-1} \\ \theta_{\text{opt}} &= \arg\min_{\theta \in \Theta} \operatorname{tr}\left[C_{\text{LMMSE}}(\theta)\right] \end{aligned} \tag{6}$$

where $C_{\text{LMMSE}}(\theta)$ quantifies the residual PCA-envelope estimation uncertainty that remains after applying a given encoding and its matched optimal linear reconstructor. $\theta_{\text{opt}}$ denotes the encoding parameters that minimize the minimum achievable linear reconstruction risk subject to all physical constraints, from which the corresponding sensing matrix $H_{\text{opt}}$ is obtained, whereas $\Theta$ denotes the set of all feasible encoding parameters. For the optimal encoding configuration $\theta_{\text{opt}}$, the matched optimal linear reconstructor can then be computed directly in closed form:

$$\begin{aligned} K(\theta_{\text{opt}}) &= \Lambda_{\text{env}} B^{\text{T}} H_{\theta_{\text{opt}}}^{\text{T}} \left(H_{\theta_{\text{opt}}} B \Lambda_{\text{env}} B^{\text{T}} H_{\theta_{\text{opt}}}^{\text{T}} + R_n\right)^{-1} = C_{\text{LMMSE}}(\theta_{\text{opt}}) B^{\text{T}} H_{\theta_{\text{opt}}}^{\text{T}} R_n^{-1} \\ W_{\text{opt}} &= B K(\theta_{\text{opt}}) \end{aligned} \tag{7}$$

where $K(\theta_{\text{opt}})$ denotes the PCA coefficient reconstructor matched to the optimized encoder, and $W_{\text{opt}}$ is the corresponding reconstruction matrix in the full spectral space. The encoder and reconstructor are jointly determined using the same source-domain PCA prior. For different HSI architectures, the physical encoding is described by the sensing matrix $H_{\theta}$ and feasible parameter set $\Theta$. The framework is illustrated in Fig. 1c, and the full derivation is provided in Supplementary Eqs. (S7)–(S14).

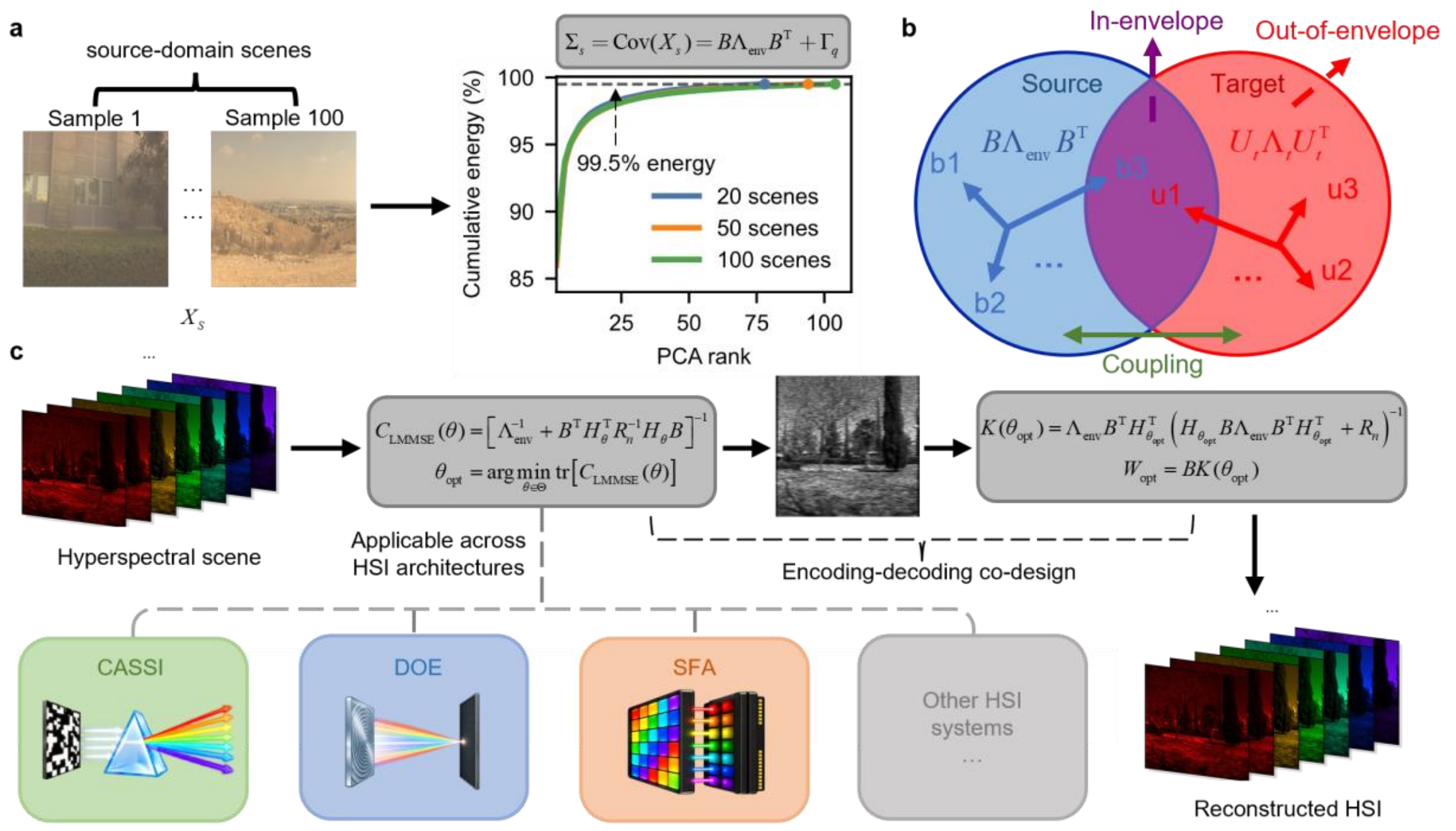

**Fig. 1 | HSI reconstruction-risk decomposition, PCA-envelope-driven encoding-decoding co-design. a,** Construction of the source-domain PCA envelope and cumulative energy as a function of PCA rank for different numbers of source-domain scenes. **b,** Relationship between the source- and target-domain statistical envelopes, highlighting the in-envelope, out-of-envelope and coupling components. **c,** PCA-envelope-driven encoding-decoding co-design across HSI architectures, including CASSI, DOE, SFA and other HSI systems.

### CASSI encoding-decoding co-design

We apply Eqs. (6) and (7) to CASSI, DOE and SFA systems with different physical encoding mechanisms. For each architecture, we construct a mapping from the physical parameters $\theta$ to the sensing operator $H_\theta$, optimize the encoding parameters within the corresponding physically realizable design space, and determine the matched linear reconstructor under the same source-domain PCA-envelope prior. Performance is then evaluated through simulations using real hyperspectral datasets. System-specific forward models and encoding-decoding details are provided in Supplementary Notes 4 and 5, and the convergence of the encoder optimization is shown in Supplementary Fig. 3.

We first apply the unified framework to CASSI. By substituting the parameterized CASSI sensing operator into Eq. (6) and minimizing the trace of the LMMSE error covariance of the PCA-envelope coefficients subject to a binary-mask constraint and a 25% open fraction, we obtain the PCA-optimized mask (Fig. 2a). We compare six configurations in Fig. 2b–d and Table 1 to distinguish the effects of encoding design, the statistical prior and the reconstructor. Ours (PCA-LMMSE with the PCA-optimized mask) represents the complete encoding-decoding design proposed here; PCA + GMRF uses LMMSE reconstruction with a spatial-spectral Gaussian Markov random field (GMRF) prior, a well-established prior in statistical image modelling; PCA + DAUHST uses five-stage DAUHST[14], a high-performance Transformer-based deep unfolding reconstructor for CASSI; and PCA + ResUNet applies gated nonlinear residual correction to the PCA-LMMSE reconstruction. All four configurations use the same PCA-optimized mask. Initial + LMMSE and Initial + DAUHST instead use the initial random mask to assess the effect of encoding optimization across different reconstructors. All configurations are evaluated on the same ten target scenes under identical light throughput and noise conditions, with a mean measurement SNR of approximately 28 dB. The average results over ten scenes are reported in Table 1, while scene-wise performance, representative spectra and colour reconstructions are shown in Fig. 2b–d. Complete full-scene reconstruction results are provided in Supplementary Fig. 5.

**Table 1** Mean reconstruction performance over ten ICVL test scenes in the CASSI experiment.

| Method | Encoder | Reconstructor | PSNR (dB) | SAM (deg) | SSIM | SNR (dB) |
|---|---|---|---|---|---|---|
| **Ours** | PCA-optimized | PCA-LMMSE | 34.76 | 2.01 | 0.9106 | 28.03 |
| PCA + GMRF | PCA-optimized | GMRF-LMMSE | 29.95 | 4.80 | 0.8945 | 28.03 |
| PCA + DAUHST | PCA-optimized | DAUHST-5stg | 35.91 | 2.16 | 0.9335 | 28.03 |
| PCA + ResUNet | PCA-optimized | PCA-LMMSE + ResUNet | 36.98 | 1.57 | 0.9459 | 28.03 |
| Initial + LMMSE | Initial random | PCA-LMMSE | 34.28 | 2.14 | 0.9061 | 28.02 |
| Initial + DAUHST | Initial random | DAUHST-5stg | 33.66 | 3.61 | 0.9190 | 28.02 |

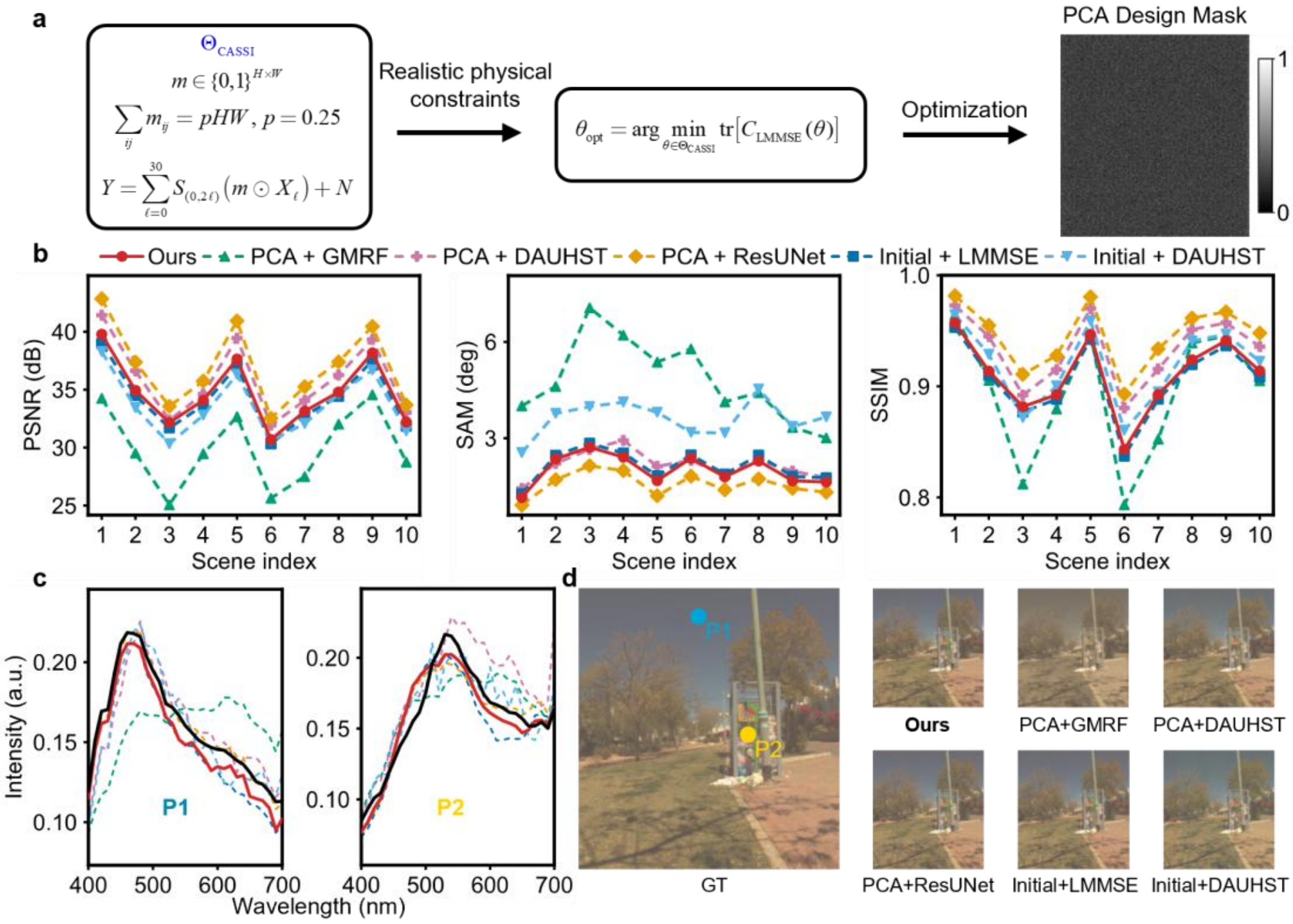

**Fig. 2 | PCA-envelope-driven CASSI encoding-decoding co-design and reconstruction performance. a,** Physically constrained optimization of the CASSI encoding mask. **b,** Scene-wise PSNR, SAM and SSIM across ten target scenes. **c,** Reconstructed and ground-truth spectra at two representative pixels (P1 and P2), whose locations are marked in the ground-truth colour image in **d**; method colours follow **b**, while the ground truth is shown as a black solid line, Ours as a red solid line and the other reconstructions as thinner dashed lines. **d,** Ground-truth colour composite with the locations of P1 and P2, together with colour reconstructions from the six configurations in **b** for Scene 4.

With the same PCA-optimized mask, Ours outperforms the Gaussian Markov baseline by 4.81 dB in PSNR, 2.78° in SAM and 0.0162 in SSIM, indicating that the PCA-envelope prior and its matched LMMSE reconstructor exploit the CASSI measurements more effectively. Ours achieves a mean PSNR only 1.15 dB lower than PCA + DAUHST, while reducing SAM by 0.15°. With the same initial mask, our linear reconstructor also achieves higher PSNR and lower SAM than DAUHST (34.28 versus 33.66 dB and 2.14° versus 3.61°), showing that its competitive performance does not arise solely from encoder optimization. Conversely, DAUHST itself benefits substantially from the PCA-optimized mask: relative to the initial mask, its PSNR increases from 33.66 to 35.91 dB (+2.25 dB), SAM decreases from 3.61° to 2.16° (−1.45°), and SSIM increases from 0.9190 to 0.9335 (+0.0145). This indicates that the optimized encoder is not limited to the matched linear reconstructor and can also improve learned nonlinear reconstruction. These comparisons show that the proposed linear reconstructor achieves competitive performance and that the optimized encoder also benefits nonlinear reconstruction.

**Sample-efficient reconstruction with limited source data**

**Table 2** Reconstruction performance and reconstructor training time under limited source-domain data.

| Reconstructor | Source scenes | PSNR (dB) | SAM (deg) | SSIM | Reconstructor training time (min) | SNR (dB) |
|---|---|---|---|---|---|---|
| **Ours** | 5 | 34.57 | 2.08 | 0.9086 | 0 | 28.09 |
| | 20 | 34.66 | 2.03 | 0.9085 | 0 | 28.04 |
| DAUHST-5stg | 5 | 25.13 | 11.55 | 0.4604 | 24.80 | 28.09 |
| | 20 | 30.43 | 5.75 | 0.8118 | 97.74 | 28.04 |

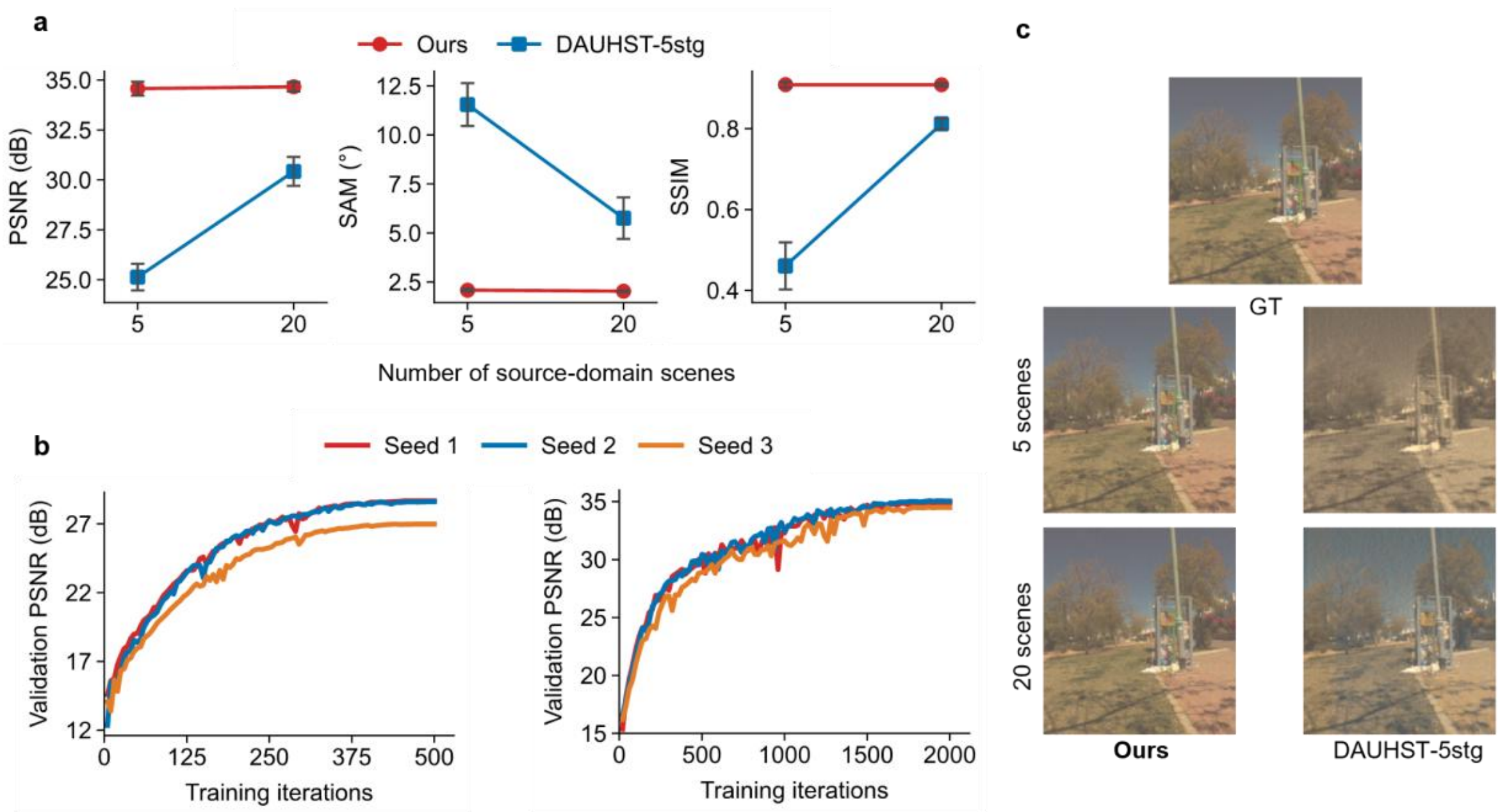

**Fig. 3 | Robust linear reconstruction with limited source-domain data. a,** Reconstruction performance versus the number of source-domain scenes. Points represent the mean across three seed runs (n = 3), and error bars indicate ±1 sample standard deviation. **b,** DAUHST-5stg validation PSNR versus training iterations using 5 (left) or 20 (right) source-domain scenes. Curves represent three seed runs with different source-scene subsets, evaluated on the same validation set. **c,** Representative colour composites for Seed 1, Scene 4.

To evaluate source-domain sample efficiency, we generated three reproducible and mutually disjoint 20-scene subsets from the complete 100-scene ICVL source pool, with a 5-scene subset nested within each 20-scene subset. For each subset, we re-estimated the PCA-envelope prior and redesigned the CASSI mask, and then compared our linear reconstructor with DAUHST-5stg trained using the same source scenes and mask on the same ten ICVL target scenes; all other settings were held fixed. Detailed experimental settings and per-seed results are provided in Supplementary Note 4.5.

With only five independent source-domain scenes, our method achieves a mean PSNR of 34.57 dB, exceeding DAUHST-5stg by 9.43 dB, while also yielding substantially lower SAM and higher SSIM (Fig. 3a,c and Table 2). As the number of source-domain scenes increases, DAUHST-5stg performs better, reducing the PSNR gap to 4.23 dB with 20 scenes and surpassing our linear reconstructor with 100 scenes. This improvement, however, comes with increasing computational cost: DAUHST-5stg requires 24.80 min of GPU training with five scenes, 97.74 min with 20 scenes and 493.70 min with 100 scenes. The validation PSNR curves for all three random subsets approach clear plateaus by the end of training (Fig. 3b), indicating that the large performance gap in the limited-data regime is not primarily caused by incomplete optimization. Our linear reconstructor is computed directly from the source-domain PCA-envelope statistics, the CASSI forward model and the noise covariance, without GPU training of the reconstructor. Its reconstruction performance changes only slightly as the number of source-domain scenes increases, showing greater sample efficiency and lower reconstructor training cost than DAUHST-5stg. This difference is consistent with the linear reconstructor relying on estimated source-domain statistics, while DAUHST learns a nonlinear mapping from the training scenes. Complete full-scene reconstruction results are provided in Supplementary Fig. 6a.

The CASSI experiments show that, although our linear reconstructor is derived from an analytical risk criterion rather than optimized for benchmark performance, it achieves, to the best of our knowledge, SOTA linear reconstruction performance for single-shot CASSI. Under controlled conditions with sufficient source-domain data, its overall performance with the PCA-optimized mask approaches that of DAUHST-5stg, while under the same initial mask it achieves higher PSNR and lower SAM. This advantage becomes more pronounced when source-domain data are limited, where our linear reconstructor significantly outperforms fully trained DAUHST-5stg in PSNR, SAM and SSIM. The linear reconstructor is computed directly from source-domain PCA-envelope statistics, the physical forward model and noise covariance, without GPU training of the reconstructor. It also provides strong reconstruction performance, sample efficiency and physical interpretability.

**Cross-domain reconstruction with broader source coverage**

To evaluate cross-domain reconstruction, we reconstructed ten fixed KAUST target scenes using PCA-envelope priors derived from either ICVL alone or a broader mixed-source dataset, under both initial and optimized CASSI encoders. The composition and preprocessing of the source-domain datasets are detailed in Supplementary Note 2.

The cross-domain experiment tests the numerical accuracy of the five-term decomposition and its ability to reveal the sources of reconstruction error (Fig. 4a,b and Table 3). Across all four source-prior and encoder configurations, the sum of the five analytical terms agrees with the empirical mean full-scene reconstruction risk within 0.2% (Fig. 4a). The individual terms further show that different error components respond differently to system redesign and statistical expansion. Under the ICVL prior with the initial encoder, encoder–reconstructor co-design reduces the in-envelope estimation term $T_{\mathrm{in}}$ from 2784.4 to 2292.8 and the noise propagation term $T_{\mathrm{noise}}$ from 757.3 to 606.2, while the dominant out-of-envelope term $T_{\mathrm{out}}$ changes only slightly from 26285.9 to 25859.0. In contrast, replacing the ICVL prior with the broader Mixed Source prior reduces $T_{\mathrm{out}}$ from 26285.9 to 10303.7 and $T_{\mathrm{mean}}$ from 3816.9 to 1408.9. These results indicate that system redesign mainly improves estimation-related errors, whereas broader statistical coverage is more effective for reducing mismatch-related errors.

**Table 3** Mean full-scene risk decomposition and reconstruction performance in the cross-domain CASSI experiment.

| Source domain | Encoder | $T_{in}$ | $T_{out}$ | $T_{cpl}$ | $T_{noise}$ | $T_{mean}$ | Analytical risk | Empirical risk | PSNR (dB) | SAM (deg) | SSIM | SNR (dB) |
|---|---|---|---|---|---|---|---|---|---|---|---|---|
| ICVL | Initial | 2784.4 | 26285.9 | −848.4 | 757.3 | 3816.9 | 32796.0 | 32834.7 | 25.87 | 6.99 | 0.7341 | 29.72 |
| | Optimized | 2292.8 | 25859.0 | −808.2 | 606.2 | 3761.9 | 31711.8 | 31711.1 | 26.13 | 6.91 | 0.7323 | 29.73 |
| Mixed | Initial | 1778.7 | 10303.7 | 270.4 | 869.6 | 1408.9 | 14631.3 | 14659.1 | 28.78 | 4.17 | 0.8038 | 29.72 |
| | Optimized | 1310.2 | 9885.1 | 192.1 | 689.8 | 1300.1 | 13377.4 | 13373.8 | 29.26 | 4.00 | 0.8121 | 29.73 |

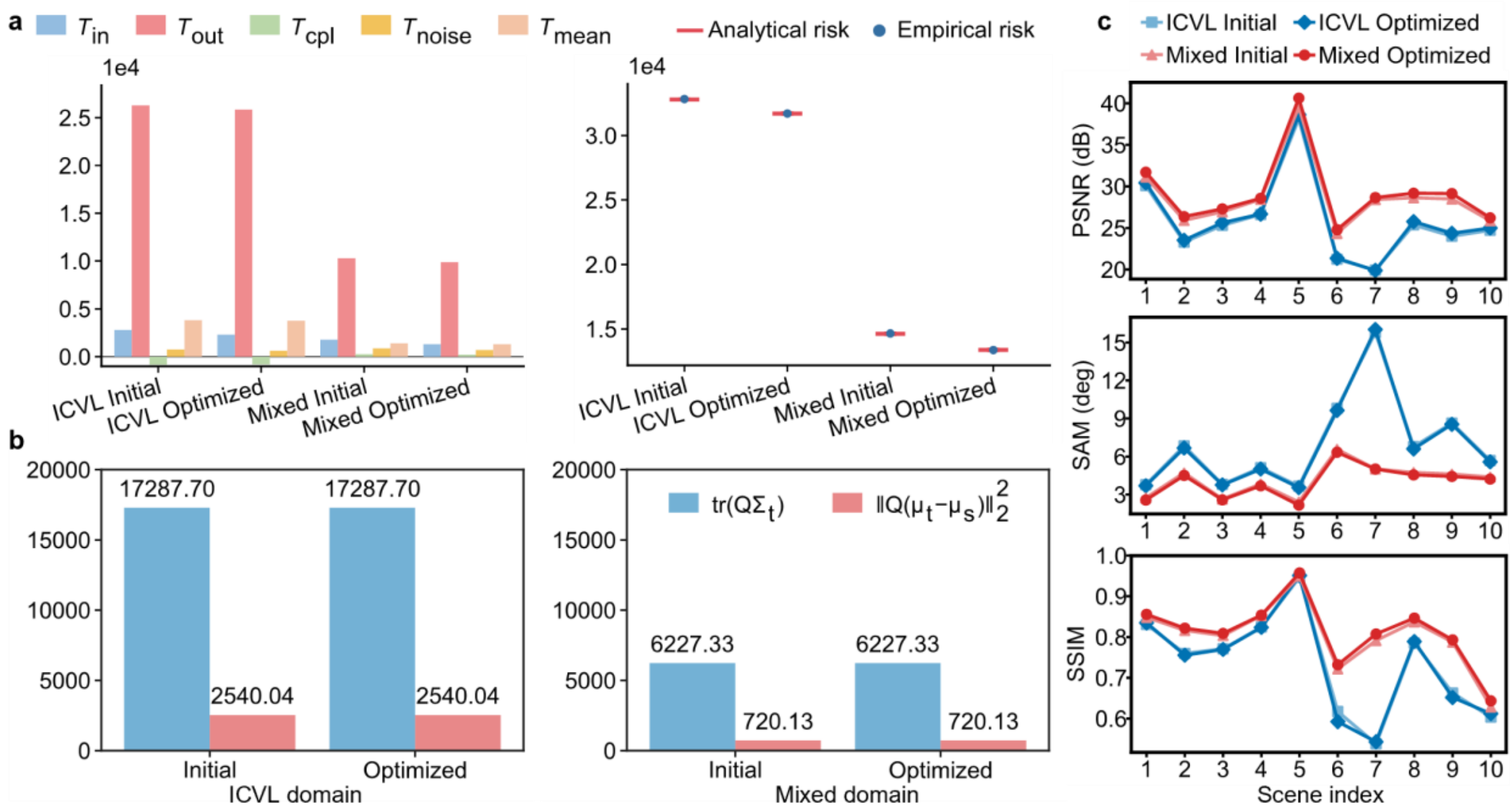

**Fig. 4 | Cross-domain CASSI reconstruction-risk analysis and reconstruction performance. a,** Five-term analytical risk decomposition and comparison with empirical risk for the four source-prior and encoder configurations. **b,** Covariance and mean-shift components of the statistical-coverage floor for the ICVL and Mixed Source priors, which remain unchanged under encoder optimization. **c,** Scene-wise PSNR, SAM and SSIM for the same four configurations.

The five-term decomposition identifies how the error components change, while Eq. (5) bounds the remaining scope for co-design under a fixed PCA envelope (Fig. 4a,b). With the ICVL prior and initial encoder, the statistical-coverage floor is 19,827.73, accounting for 60.5% of the analytical reconstruction risk. Thus, at most 39.5% of the current risk can be reduced through further co-design without changing the PCA envelope. Replacing the ICVL prior with the patch-count-matched Mixed Source prior reduces this floor to 6,947.46, a decrease of 65.0%. For either source prior, the floor remains unchanged after encoder optimization (Fig. 4b). Consistent with the error decomposition, co-design reduces the empirical risk by only 3.4% under the ICVL prior, whereas broader source coverage reduces it by 55.4%. With the Mixed Source prior, co-design provides a further 8.8% reduction. The decomposition and risk bound therefore provide complementary information for deciding whether to improve statistical coverage or further optimize the imaging system.

The scene-wise reconstruction metrics show the same general trend (Fig. 4c): configurations with lower empirical risk usually have higher PSNR and SSIM and lower SAM. Broader statistical coverage leads to the largest improvement, while the additional gains from co-design are smaller once the statistical mismatch has been reduced. Under the ICVL prior, co-design changes the reconstruction metrics only slightly, consistent with the limited reduction in empirical risk. Complete full-scene reconstruction results are provided in Supplementary Fig. 6b.

**DOE encoding-decoding co-design**

**Table 4** Mean reconstruction performance in the DOE experiment.

| Method | Encoder | Reconstructor | PSNR (dB) | SAM (deg) | SSIM | PSF energy loss (%) | SNR (dB) |
|---|---|---|---|---|---|---|---|
| Initial DOE | Initial structure | PCA-LMMSE | 30.76 | 3.17 | 0.8659 | 3.42 | 29.44 |
| PCA Design | PCA-optimized | PCA-LMMSE | 33.40 | 3.00 | 0.9087 | 4.29 | 29.06 |

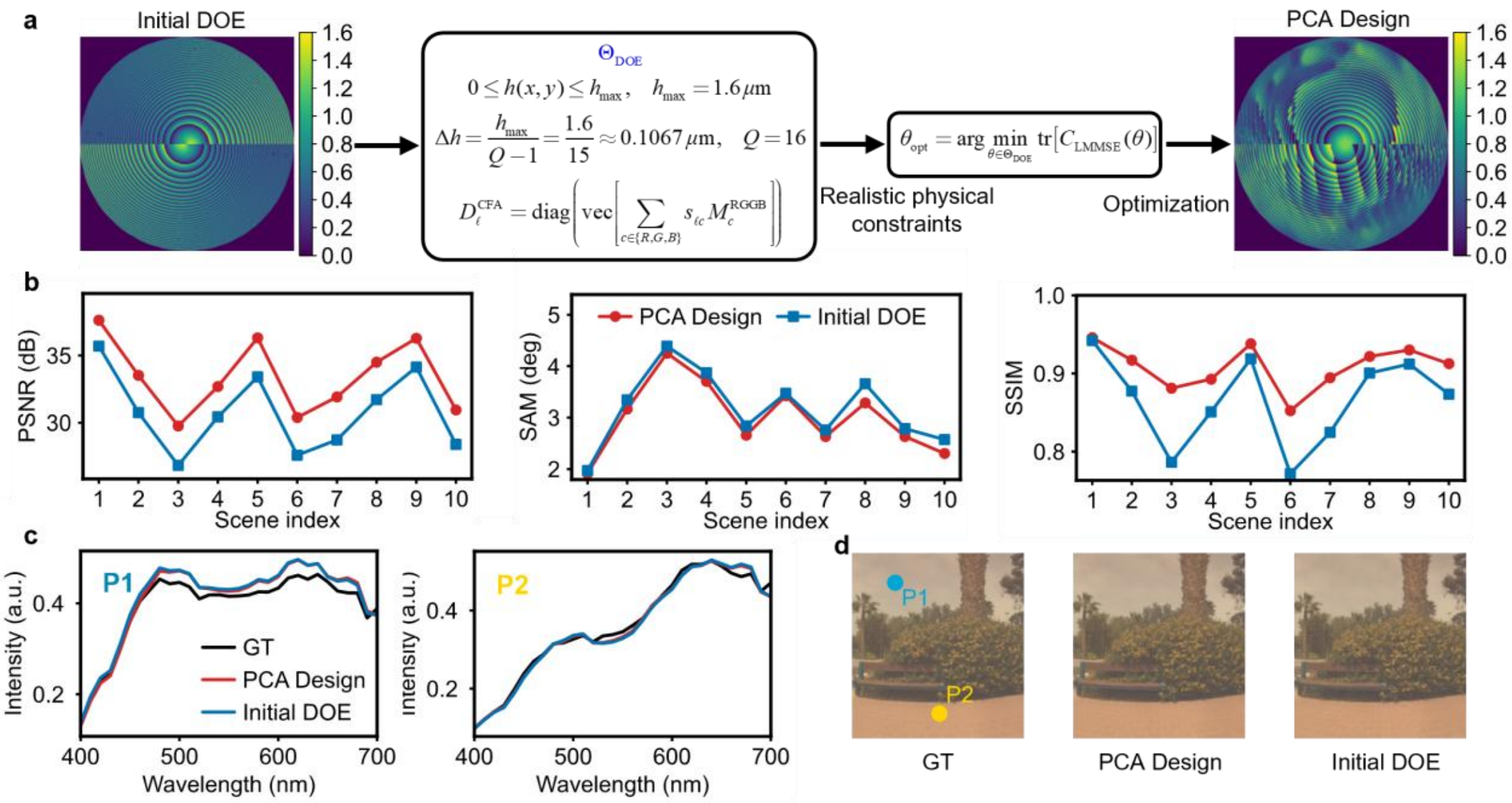

**Fig. 5 | PCA-envelope-driven DOE optimization and hyperspectral reconstruction. a,** Physically constrained DOE optimization from the Initial DOE to the PCA Design DOE. **b,** Scene-wise PSNR, SAM and SSIM across ten target scenes. **c,** Reconstructed and ground-truth spectra at two representative pixels (P1 and P2), whose locations are marked in the ground-truth colour image in d. **d,** Ground-truth colour composite with the locations of P1 and P2, together with colour reconstructions from PCA Design and Initial DOE.

To test the framework in a different imaging architecture, we apply it to a diffractive HSI system comprising a multilevel DOE and a Bayer-pattern detector (Canon 1D Mark III)[15]. The wavelength-dependent DOE response is incorporated into the parameterized sensing operator, and the DOE height profile is optimized within the physically realizable design space using the same PCA-envelope-driven LMMSE criterion. Optimization of the Initial DOE produces the PCA Design shown in Fig. 5a. The corresponding wavelength-dependent PSFs before and after PCA optimization are shown in Supplementary Fig. 2.

As shown in Fig. 5b and Table 4, PCA Design increases the mean PSNR from 30.76 to 33.40 dB, reduces SAM from 3.17° to 3.00°, and improves SSIM from 0.8659 to 0.9087. The spectra at P1 and P2 are shown in Fig. 5c. PCA Design follows the ground truth more closely at P1, while both designs give similar spectra at P2. Representative colour reconstructions are shown in Fig. 5d. These gains are obtained despite a slightly lower measurement SNR (29.06 versus 29.44 dB), indicating that the improvement arises from more informative spectral encoding rather than increased measurement signal power. These results show that the PCA-envelope-driven encoding–decoding co-design method can also be applied to diffractive HSI systems with different physical encoding mechanisms. Full-scene reconstruction results are provided in Supplementary Fig. 7a.

**SFA encoding-decoding co-design**

**Table 5** Mean reconstruction performance in the SFA experiment.

| Method | Encoder | Reconstructor | SNR (dB) | PSNR (dB) | SAM (deg) | SSIM |
|---|---|---|---|---|---|---|
| Initial Design | Initial random | PCA-LMMSE | 27.14 | 37.16 | 2.51 | 0.9597 |
| PCA Design | PCA-optimized | PCA-LMMSE | 29.30 | 38.70 | 1.78 | 0.9621 |

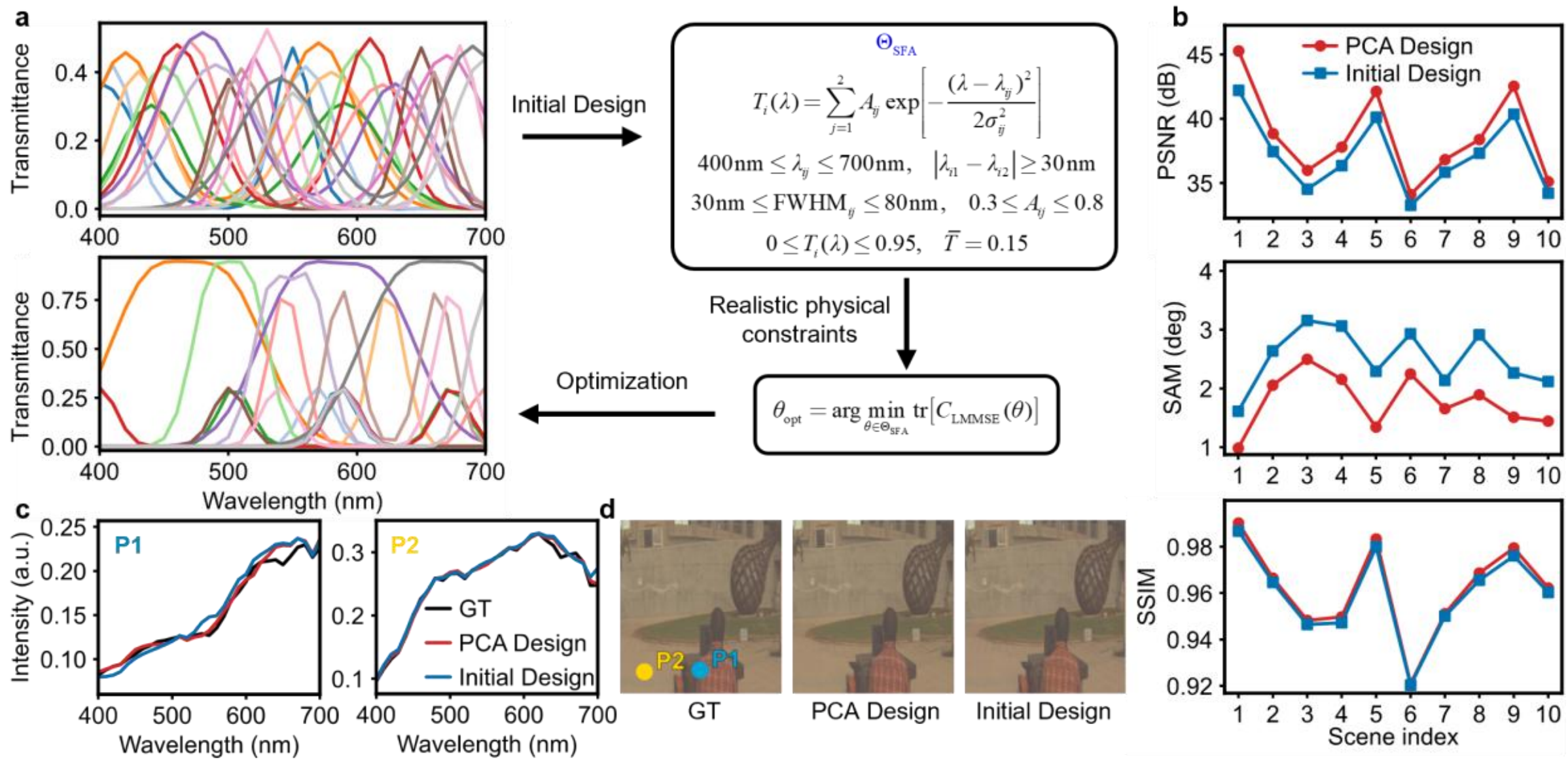

**Fig. 6 | PCA-envelope-driven SFA encoding-decoding co-design and reconstruction performance. a,** Initial and optimized spectral responses of the 16 SFA filters; colours denote individual filters. **b,** Scene-wise PSNR, SAM and SSIM across ten target scenes. **c,** Reconstructed and ground-truth spectra at two representative pixels (P1 and P2), whose locations are marked in the ground-truth colour image in **d**. **d,** Ground-truth colour composite with the locations of P1 and P2, together with colour reconstructions from PCA Design and Initial Design.

We also apply the design framework in Eqs. (6) and (7) to the SFA system. The SFA consists of 16 spectral filters arranged in a fixed 4×4 periodic mosaic, with each filter integrating the incident spectrum according to its transmission response. Fig. 6a shows the initial and optimized filter responses under the physical constraints. Initial Design provides broad spectral coverage, whereas PCA Design adjusts the peak positions, bandwidths and amplitudes of the 16 filters to redistribute spectral coverage and inter-band overlap.

Fig. 6b and Table 5 show that PCA Design improves the mean PSNR by 1.54 dB, reduces SAM by 0.73°, and increases SSIM by 0.0023 relative to Initial Design. Under the same average transmittance and noise constraints, PCA Design also increases the measurement SNR from 27.14 to 29.30 dB because the coupling between the filter responses and scene spectra alters the effective signal power. The spectra at P1 and P2 are shown in Fig. 6c, while representative colour reconstructions are shown in Fig. 6d. Therefore, the source-domain PCA envelope can guide filter-response-LMMSE co-design to improve spectral coverage, complementarity and reconstruction accuracy under a fixed transmittance budget. Full-scene reconstruction results are provided in Supplementary Fig. 7b.

## Discussion

We present a five-term reconstruction-risk decomposition for linear hyperspectral imaging and a PCA-based method for joint encoding and reconstruction. The decomposition identifies the contributions of different error sources, while the statistical-coverage lower bound for PCA-envelope-constrained reconstruction helps assess the potential for further system optimization. Experiments with CASSI, DOE and SFA show that the co-design method can be applied to different physical encoding architectures. The cross-domain CASSI results further show how encoder optimization and broader source coverage affect reconstruction risk. The matched linear reconstructor performs well with only a few source scenes and requires no GPU training, reducing the data requirements and reconstructor training cost. These results suggest that source-data coverage should be considered alongside physical sensing and reconstruction in HSI system design. The approach may also apply to other computational imaging systems with linear sensing models. Future work may consider richer statistical priors, more realistic noise models and nonlinear or learned reconstruction. The study also involved collaboration between researchers and an autonomous AI research system, an approach that may be useful in related computational imaging research.

## Methods

### Datasets and preprocessing

The main-text experiments use the ICVL[16] and KAUST[17] datasets as source or target domains, whereas the Supplementary Information further uses the KAIST[18] dataset to verify the agreement between the five-term risk decomposition and Monte Carlo mean-squared error. Dataset-specific normalization is described in Supplementary Note 2.2 and defined in Supplementary Eq. (S16).

### Patch-based PCA-envelope estimation

A full hyperspectral scene can contain tens of millions of spatial-spectral variables, whereas only a limited number of source-domain scenes are typically available, making full-scene covariance estimation computationally expensive and statistically unstable. We therefore estimate the source-domain PCA envelope from local spatial-spectral patches extracted from source-domain scenes, reducing covariance dimensionality while increasing the number of samples available for estimation. Each patch is vectorized as defined in Supplementary Eq. (S15). Details of dataset patch extraction, source-target scene separation and PCA-envelope estimation are provided in Supplementary Note 2.1, whereas the embedding of the resulting PCA means and bases into full-scene sensing and reconstruction is described in Supplementary Note 4.1.

### Linear measurement and noise model

For the parameterized linear measurement model in Eq. (1), we take the sensing operator to specify the conditional mean of the detector output:

$$\mathbb{E}[y|\ x] = H_\theta x + b \tag{8}$$

For real imaging systems operating in the linear, non-saturating regime, with the system response and bias adequately calibrated, Eq. (8) holds to within experimental accuracy. All theoretical derivations assume bias-corrected measurements, such that (b=0). Because measurement noise in practical imaging systems typically has finite second moments, its covariance matrix is well defined; to enable the inverse-covariance LMMSE formulation, we further assume it to be positive definite:

$$R_n = \mathrm{Cov}(n) \succ 0 \tag{9}$$

Optical detection noise can be approximated as Gaussian under high photon flux, and additive white Gaussian noise is widely adopted in computational imaging[19]. To isolate the effects of encoding design and prior selection under controlled and reproducible conditions, all simulations use additive white Gaussian noise satisfying the covariance condition in Eq. (9):

$$n \sim \mathcal{N}(0, \sigma_n^2 I) \tag{10}$$

**Simulation settings and controlled comparisons**

Within each controlled comparison, all non-targeted factors—including the target scenes, forward-model formulation, physical constraints, measurement or light-throughput budget, noise conditions, data normalization and evaluation metrics—are held fixed; only the designated encoder, source-domain prior or reconstruction component is varied as specified.

Specifically, the CASSI encoding experiment fixes the mask open fraction at 0.25 and the dispersion step between adjacent spectral bands at two detector pixels, with $\sigma_n = 0.08$; the cross-domain CASSI experiment fixes the number of source-domain patches and sets $\sigma_n = 0.07$ (patch settings are provided in Supplementary Note 2.3); the DOE experiment fixes the propagation model, parameterization, RGB camera response and reconstruction pipeline, with $\sigma_n = 0.025$; and the SFA experiment uses a dual-peak parameterization with centre wavelengths of 400–700 nm, a minimum peak separation of 30 nm, full widths at half maximum of 30–80 nm, peak amplitudes of 0.3–0.8, a maximum transmittance of 0.95, a fixed average transmittance of 0.15, and $\sigma_n = 0.055$. The encoder-optimization settings and total experiment runtimes are summarized in Supplementary Table 3.

To ensure consistent quantitative evaluation across the CASSI, DOE and SFA experiments, the same definitions were used for measurement signal-to-noise ratio (SNR), reconstruction peak signal-to-noise ratio (PSNR), spectral angle mapper (SAM) and structural similarity index (SSIM). Before metric evaluation, each reconstructed HSI cube was clipped to [0, 1], whereas the corresponding normalized ground-truth cube was left unchanged. For each target scene, measurement SNR was calculated in decibels from the energy ratio between the noise-free measurement and the realized additive noise; reconstruction PSNR was computed from the mean-squared error over the complete 31-band HSI cube; SAM was obtained by averaging the 31-band spectral angles over all spatial pixels; and SSIM was evaluated independently for each spectral band and then averaged over all 31 bands. The reported values are arithmetic means of the corresponding scene-wise metrics across the ten target scenes.

**Initial DOE structure**

The DOE height profile is initialized using a dual-period rotational focusing structure in which the design wavelength varies periodically with azimuth:

$$h_{\mathrm{init}}(x,y) = Q_{16}\left\{ A(x,y) \frac{\mathrm{mod}[-\mathrm{OPD}(x,y), \lambda]}{n(\lambda)-1} \right\} \tag{11}$$

where $\mathbf{Q}_{16}$ denotes the 16-level height-quantization operator, A(x,y) is the DOE aperture function, equal to 1 within the effective aperture and 0 outside, and $\mathrm{OPD}(x,y)$ is the optical path difference required for focusing. The initial focal length is 200 mm, $n(\lambda)$ accounts for material dispersion, and fused silica is used as the DOE material. The DOE sampling pitch is 5 μm and its size is 5 mm. The sensor comprises 3888×2592 pixels with a pixel pitch of 7.2 μm.

PCA-envelope-driven DOE optimization minimizes the normalized trace of the LMMSE error covariance, with auxiliary regularization terms for quantization, height total variation, and mean height weighted by $10^{-3}$, $10^{-4}$ and $10^{-3}$, respectively. These terms are used only for DOE design and are not included in the matched LMMSE reconstruction.

**Researcher–AI collaborative theoretical exploration**

The initial theoretical exploration was supported by an autonomous AI research system (Qiushi Engine). The researchers defined the scientific questions and evaluated the results, while the AI system explored alternative approaches, examined assumptions and developed candidate derivations. During this process, the system found that improved optical measurements did not always provide information that the reconstruction algorithm could effectively use. This finding prompted a shift from mask-only optimization to encoding–reconstruction co-design based on a common statistical envelope. The researchers then verified, corrected, generalized and interpreted the candidate results. Across the two studies, the AI system made 3,332 model calls and 1,209 tool calls, processing approximately 199.9 million input tokens and 5.5 million output tokens. Further details are provided in Supplementary Note 7 and Supplementary Table 5.

**Acknowledgments**

The work at Zhejiang University was sponsored by the Science Challenge Project No. TZ2025015 (Y.Y.), by Key Research and Development Program of the Ministry of Science and Technology under Grant No. 2022YFA1404900 (Y.Y.), 2022YFA1405200 (Y.Y.), 2022YFA1404704 (H.C.), and 2022YFA1405201 (H.C.), by the National Natural Science Foundation of China (NSFC) under Grants U25D8017 (Y.Y.), U25A20520 (H.C.), 62475228 (H.C.), by the Class D for Young Scientific Research Peak Creation No. K20250230 (Y.Y.), by the Fundamental Research Funds for the Zhejiang Provincial Universities No. 226-2025-00231 (Y.Y.), by the China Postdoctoral Science Foundation No. 2025M780537 (F.C.)，and 2026T190149 (F. C.).

**Author contributions**

Y.Y. conceived the original idea. S.H. and S.Y. validated the hypotheses, extended the theoretical framework, and designed the numerical simulations for validation. Z.Y., F.C. and Y.Y. supervised the project. S.H. wrote the original draft with input from S.Y., T.Y., Z.Y., F.C., H.C. and Y.Y. All authors reviewed and approved the final manuscript.

**Competing interests**

The authors declare no competing interests.

**Data availability**

The data supporting the findings of this study are available within the Article and Supplementary Information. Additional data and information are available from the corresponding authors upon reasonable request.